\documentclass[useAMS,usenatbib]{mn2e}
\usepackage{graphicx}
\title[The very massive binary NGC3603-A1]{The very massive binary NGC3603-A1}

\author[O. Schnurr et al.]{O. Schnurr$^{1,2}$\thanks{E-mail:
o.schnurr@sheffield.ac.uk}, A. F. J. Moffat$^{1}$, N. St-Louis$^{1}$,
J. Casoli$^{1,3}$, \newauthor and A.-N. Chen\'e$^{1,4}$
\vspace{3mm}\\ $^{1}$D\'ept. de Physique , Universit\'e de Montr\'eal,
C. P. 6128, succ. centre-ville, Montr\'eal (Qc) H3C 3J7, and Centre de Recherche\\en
Astrophysique du Qu\'ebec, Canada\\
$^{2}$Dept. of Physics and Astronomy, University of Sheffield, Hicks
Building Hounsfield Road, Sheffield S3 7RH, United Kingdom\\
$^{3}$Ecole Normale Sup\'erieure, 45, rue d'Ulm, 75230 Paris C\'edex 05, France\\
$^{4}$Herzberg Institute of Astrophysics, 5071 West Saanich Road,
Victoria (BC) V9E 2E7, Canada}

\begin{document}

\date{Version 17 June 2008}

\pagerange{\pageref{firstpage}--\pageref{lastpage}} \pubyear{2008}

\maketitle

\label{firstpage}

\begin{abstract}
Using VLT/SINFONI, we have obtained repeated AO-assisted, NIR
spectroscopy of the three central WN6ha stars in the core of the very
young ($\sim1$ Myr), massive and dense Galactic cluster NGC3603. One
of these stars, NGC3603-A1, is a known 3.77-day, double-eclipsing
binary, while another one, NGC3603-C, is one of the brightest X-ray
sources among all known Galactic WR stars, which usually is a strong
indication for binarity. Our study reveals that star C is indeed an
8.9-day binary, although only the WN6ha component is visible in our
spectra; therefore we temporarily classify star C as an SB1
system. A1, on the other hand, is found to consist of two
emission-line stars of similar, but not necessarily of identical
spectral type, which can be followed over most the orbit. Using radial
velocities for both components and the previously known inclination
angle of the system, we are able to derive absolute masses for both
stars in A1. We find $M_{\rm 1} = (116 \pm 31) M_{\sun}$ for the primary
and $M_{\rm 2} = (89 \pm 16) M_{\sun}$ for the secondary component of
A1. While uncertainties are large, A1 is intrinsically half a
magnitude brighter than WR20a, the current record holder with 83 and
82 $M_{\sun}$, respectively; therefore, it is likely that the primary
in A1 is indeed the most massive star weighed so far.
\end{abstract}


\section{Introduction}

While models maintain that in the early Universe, the first-generation
of stars were very massive and reached masses between 100 and 1000
M$_{\odot}$ (e.g. \citealt{NakaUme01}; \citealt{Schaerer02}), it is
generally accepted that under present-day conditions, relatively fewer
massive stars are formed, i.e. that the initial-mass function (IMF) is
much steeper and, more importantly, has a cut-off occurring around 150
M$_{\odot}$ (\citealt{WeidKroup04}; \citealt{Figer05}). So far,
however, whenever Keplerian orbits of binary systems are used to weigh
stars -- the only way to obtain reliable, least model-dependent masses
--, measured masses fall short by almost a factor of two with respect
to the putative cut-off. Currently, stars with the highest known
masses are both WN6ha components of the Galactic WR binary WR20a, with
83 and 82 M$_{\odot}$, respectively (\citealt{Rauw04};
\citealt{Bonanos04}), and the O3f/WN6 star in the Galactic binary
WR21a with a minimum mass of 87 M$_{\odot}$
(\citealt{Gamen08pp}). Significantly more massive stars, however, have
so far remained elusive.

A most remarkable result that has emerged from the search for very
massive stars is, however, that the highest Keplerian masses are
\emph{not} found among absorption-line O-type stars -- rather, masses
of these stars remain below $\sim60$ M$_{\odot}$
(e.g. \citealt{Lamontagne96}; \citealt{Massey02}) --, but among
Wolf-Rayet (WR) stars, more precisely among the so-called WN5-7ha (or
WN5-7h) stars, an extremely luminous and hydrogen-rich subtype of the
nitrogen-sequence WR stars. Contrary to classical WR stars, which are
identified with bare, helium-burning cores of evolved massive stars,
theoretical work confirms that these luminous WN5-7ha stars are
hydrogen-burning, unevolved objects (\citealt{deKoter97};
\citealt{CroDess98}) which mimic the spectral appearance of WR stars
because their high luminosities drive dense and fast winds
(\citealt{GraeHam08pp}).

If it is true that the most massive stars can be found in the most
massive clusters (\citealt{Bate02}; \citealt{WeidKroup06}), then it
can be expected that the most massive WN5-7ha stars are hosted by the
most massive among the youngest, least evolved clusters known. NGC3603
is a Galactic example of such a cluster, and virtually a clone of its
more famous LMC counterpart, the supermassive cluster R136, at the
core of the giant HII region 30 Dor ({\citealt{Moff94}). NGC3603's
very core, itself denoted HD 97950, contains three extremely luminous
WN6ha stars (\citealt{Drissen95}), with stellar luminosities well in
excess of $10^{6}$ L$_{\odot}$ (\citealt{deKoter97};
\citealt{CroDess98}). \citet{MoffNiem84} found from radial-velocity
variations in unresolved spectra of HD 97950 that one of these WN6ha
stars must be a binary with a period of 3.772 days, which was
confirmed by \citet{Moff85}. In a recent study, \citet{Moff04} used
HST-NICMOS J-band photometry to confirm that A1 is indeed a
double-eclipsing binary with that period, while the other WN6ha stars
B and C did not vary above the $\sim0.05$ mag noise level. However,
star C shows an extremely large X-ray luminosity ($L_{\rm x} =$
several $10^{34}$ ergs$^{-1}$; \citealt{Moff02}), which is a strong
indication of binarity given that colliding winds in binary systems
generate copious amounts of hard X-rays (e.g. \citealt{Usov92}).

The double-eclipsing nature of A1 offers the rare opportunity to
directly measure the mass of an extremely luminous binary system by
simple, least model-dependent Keplerian orbits. Here we report the
results of a spectroscopic monitoring campaign of the three central
WN6ha stars in NGC3603. This Letter is organized as follows: In Section
2, we briefly describe the observations and data reduction. In Section
3, data analysis is described and results are presented. Section 4
summarizes the Letter.

\section{Observations and data reduction}

Observations were carried out in service mode at the VLT-UT4 under
Program-ID P75-0576.D. We obtained repeated K-band (1.95 to 2.45
$\mu$m) spectroscopy using SINFONI (\citealt{Eisenhauer03};
\citealt{Bonnet04}) with adaptive-optics (AO) correction to obtain the
highest possible spatial and spectral resolution. The field of view
was $0.8'' \times 0.8''$ with a ``spaxel'' scale of 12.5 mas $\times$
25 mas. Each of our three target WN6ha stars was observed
individually.

Total exposure times were 84s per star and per visit, each organized
in 4 detector integration times (DITs) of 21s. Given the brightness of
our targets ($K \sim$ 7--8mag) and the AO deployment, no dedicated sky
frames were taken. Other calibrations (dark and flatfield frames, and
the telluric standard star) were provided by the ESO baseline
calibration.

For most of the data reduction steps, ESO's pipeline was used
(cf. \citealt{Abuter06}). Standard reduction steps were taken. The
two-dimensional spectra produced by each illuminated slitlet were
individually extracted using IRAF, and combined into one
wavelength-calibrated spectrum per star and visit. A main-sequence
B-type star was used for telluric corrections. To remove the B star's
Br$\gamma$ absorption which coincides with the Br$\gamma$/He\textsc{i}
$\lambda$2.166 $\mu$m emission blend of the target WN6ha stars, a
Lorentzian was fitted to the absorption line and subtracted from the B
star's spectrum. Residuals were very small, and proved to be harmless
in the subsequent analysis. Finally, science spectra were rectified by
fitting a low-order spline function to the stellar continuum. The
final uniform stepwidth of the spectra was 2.45 \AA/pixel, resulting
in a conservative three-pixel resolving power of $\sim$3000, and a
velocity dispersion of $\sim$33 kms$^{-1}$/pixel.

\section{Data analysis and results}

\begin{figure}
\includegraphics[width=60mm,angle=-90,trim= 0 15 10 38,clip]{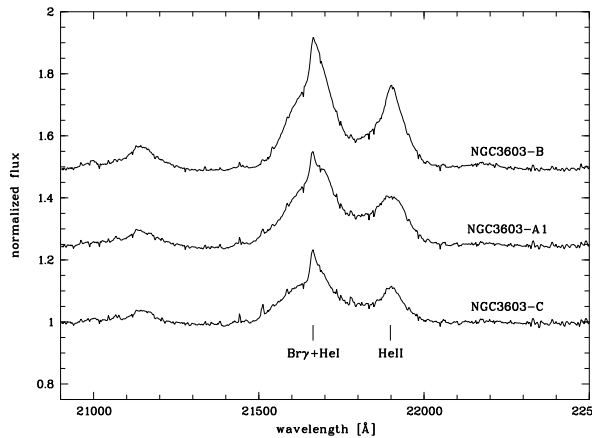}
\caption{Montage of average spectra of our three target stars, limited
to the useful spectral region. The two strongest emission lines,
Br$\gamma$/He\textsc{i} $\lambda$2.166 $\mu$m and He\textsc{ii}
$\lambda$2.188 $\mu$m are indicated. For clarity, the upper two spectra
have been shifted by 0.25 and 0.5 flux units, respectively.}
\label{montage}
\end{figure}

\begin{figure}
\includegraphics[width=70mm,angle=-90,trim= 0 10 10 150,clip]{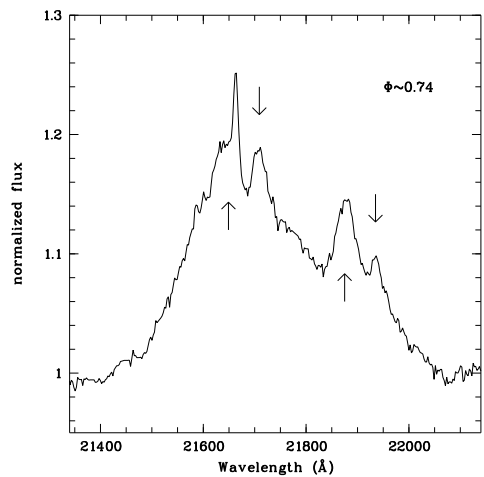}
\caption{Average of two spectra of A1, taken almost at quadrature
($\phi \sim 0.74$). Both Br$\gamma$/He\textsc{i} $\lambda$2.166 $\mu$m
and He\textsc{ii} $\lambda$2.188 $\mu$m are clearly separated; the
(stronger-lined) primary is indicated by upward arrows, while the
(slightly weaker-lined) secondary is indicated by downward arrows.}
\label{A1dedoub}
\end{figure}

Averaged spectra of the three WN6ha stars are shown in Figure
\ref{montage}. For star B, radial velocities (RVs) were measured by
cross-correlation of the region 21340 to 22140 \AA, comprising the two
emission lines Br$\gamma$/He\textsc{i} $\lambda 2.166 \mu$m and
He\textsc{ii} $\lambda 2.188 \mu$m. Star B proved to show constant RVs
over the observed timescales, with a scatter of $\sigma_{\rm RV} \sim
20$ kms$^{-1}$, which was adopted as \emph{a-posteriori} error (see
Figure \ref{orbits}). Fitting a single Gaussian to the He\textsc{ii}
$\lambda 2.188 \mu$m emission line yielded a slightly larger scatter,
$\sigma = 26$ kms$^{-1}$, but was used to obtain the systemic
velocity of the star, $\gamma = (167 \pm 6) $ kms$^{-1}$. Given is the
error of the mean, i.e. $\sigma/\sqrt(N)$, where $N=18$ is the number
of data points.

A1 was found to consist of two emission-line stars, very similar to
WR20a (\citealt{Rauw04}). An average spectrum around quadrature ($\phi
\sim 0.74$) is shown in Figure \ref{A1dedoub}. While both binary
components show Br$\gamma$/He\textsc{i} $\lambda 2.166 \mu$m and
He\textsc{ii} $\lambda 2.188 \mu$m in emission, the two components do
not show the same line strengths. In the following, we will hence
refer to the stronger-lined star as ``primary'', and to the
weaker-lined star as ``secondary''.


RVs for star C and for both components of A1 were obtained by
independently fitting Gaussians to the respective He\textsc{ii}
$\lambda 2.188 \mu$m emission peaks of the two components, because
this line is most likely less affected by wind-wind collision in the
binary than Br$\gamma$/He\textsc{i}. To take into account that in the
case of A1, the He\textsc{ii} emission peak of the secondary moves up
and down the flank of the stronger emission of the primary, a sloping,
linear continuum was simultaneously fitted. Unfortunately, fitting the
secondary's emission peaks in A1 is only possible at phases were the
respective He\textsc{ii} lines are well separated, and particularly
difficult when the weak He\textsc{ii} line of the secondary blends
with the strong Br$\gamma$/He\textsc{i} line of the primary,
i.e. during most of the ``blue'' half-wave of the companion's
motion. Moreover, due to blending with the steep slope of the
Br$\gamma$/He\textsc{i} emission, the profile of the secondary's
He\textsc{ii} emission line gets significantly distorted, thereby
further increasing the error of the fit. Both phenomena strongly
affect the determination of the secondary's orbit and hence the mass
determination of the two stars (see below). RVs for both components of
A1 are listed in Table \ref{rvA1}, and RVs for the primary component
of C are listed in Table \ref{rvC}.

\begin{table}
\caption{Journal of RVs for both components in A1 as obtained by
Gaussian fits to the He\textsc{ii} $\lambda 2.188 \mu$m emission
lines. The second column of each table provides the orbital phase
corresponding to a period of $P=3.7724$ days (from \citealt{Moff04})
and, as zero point in phase, the time of inferior conjunction $E_{0} =
2,453,765.75$ (this work). Note that it was not always possible to
measure the RVs for the secondary; see text for more details.}
\label{rvA1}
\begin{tabular}{ccrrrr}
\hline
HJD           &  $\phi$  & $RV_{1}$   & $\sigma_{1}$ & $RV_{2}$   & $\sigma_{2}$ \\
2,450,000.5+&             & \multicolumn{4}{c}{kms$^{-1}$} \\
\hline
3468.003  &  0.206  &  475  &   7   & -121  &  21   \\
3470.030  &  0.743  & -138  &  16   &  601  &   7   \\
3471.060  &  0.016  &  262  &   7   &  262  &   9   \\
3472.019  &  0.270  &  478  &   4   & -267  &  55   \\
3473.009  &  0.533  &    6  &   4   &       &   4   \\
3474.146  &  0.834  & -160  &   7   &  427  &   7   \\
3489.970  &  0.029  &  305  &   7   &  305  &   8   \\
3528.976  &  0.369  &  410  &   8   &       &       \\
3529.978  &  0.635  &  -45  &  10   &  531  &  55   \\
3531.040  &  0.916  &   48  &  10   &       &       \\
3532.961  &  0.425  &  236  &  14   &  236  &  14   \\
3552.963  &  0.727  & -184  &   7   &  630  &   7   \\
3721.321  &  0.356  &  426  &   7   &   21  &  27   \\
3723.325  &  0.888  & -121  &  27   &  210  &  21   \\
3754.240  &  0.083  &  281  &   7   &       &       \\
3758.243  &  0.144  &  285  &  21   &       &       \\
3763.292  &  0.482  &   55  &  14   &  277  &   7   \\
3765.213  &  0.991  &  177  &   7   &  177  &   8   \\
3766.362  &  0.296  &  485  &  21   &  -40  &  41   \\
3790.163  &  0.605  &   19  &  14   &       &       \\
3794.101  &  0.649  & -125  &   7   &  499  &  21   \\
3794.212  &  0.679  & -100  &  21   &  507  &  21   \\
\hline
\end{tabular}
\end{table}

\begin{table}
\caption{As for Table \ref{rvA1}, but for star C, using the period of
$P=8.89$ days and, as zero point for the phase, the time of periastron
passage $T_{0} = 2,453,547.11$ (this work).}
\label{rvC}
\begin{tabular}{ccrr}
\hline
HJD            &  $\phi$ &  $RV$       & $\sigma_{\rm RV}$ \\
2,450,000.5+   &         &\multicolumn{2}{c}{kms$^{-1}$} \\
\hline
3467.994  &  0.157  &  394  &   4    \\
3468.003  &  0.158  &  425  &   8    \\
3470.021  &  0.385  &  255  &   5    \\
3471.060  &  0.502  &   93  &   7    \\
3472.019  &  0.609  &  114  &   7    \\
3473.009  &  0.721  &   -5  &  11    \\
3474.146  &  0.849  &    0  &  10    \\
3489.970  &  0.629  &  108  &   8    \\
3528.976  &  0.016  &  263  &  10    \\
3529.978  &  0.129  &  401  &   4    \\
3531.040  &  0.249  &  381  &  10    \\
3532.961  &  0.465  &  233  &   4    \\
3552.963  &  0.715  &   19  &   7    \\
3721.321  &  0.652  &   52  &   4    \\
3723.325  &  0.878  &   -5  &  11    \\
3754.240  &  0.355  &  216  &   4    \\
3758.243  &  0.806  &  -21  &   8    \\
3759.369  &  0.932  &  179  &  14    \\
3763.292  &  0.374  &  244  &  14    \\
3765.213  &  0.590  &   89  &  27    \\
3766.353  &  0.718  &   45  &  11    \\
3766.362  &  0.719  &   34  &  11    \\
3769.387  &  0.059  &  293  &  12    \\
3790.164  &  0.396  &  255  &  11    \\
3791.102  &  0.502  &  110  &  18    \\
3794.101  &  0.839  &  -27  &  14    \\
3794.212  &  0.852  &   21  &  11    \\
\hline
\end{tabular}
\end{table}

For star C, RVs were subjected to a period search based on the
C\textsc{lean} algorithm and Scargle periodograms (see \citealt{S08}
for details). The search yielded a value of $P = 8.89 \pm 0.01$ days
(see Figure \ref{periodograms}).

\begin{figure}
\includegraphics[width=55mm,angle=-90,trim= 10 10 20 0,clip]{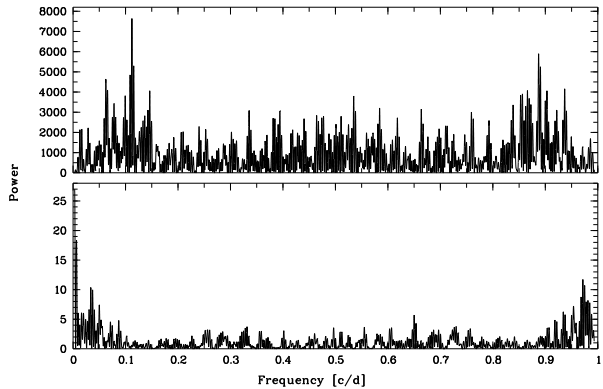}
\caption{Power spectrum (upper panel) and spectral window (lower
panel) for star C. Frequencies were searched from 0.1 to 1 d${-1}$,
corresponding to periods between 1 and 10 days. The strongest
frequency peak at $\nu \sim0.112$ d$^{-1}$ corresponds to the adopted
period of 8.89 days.}
\label{periodograms}
\end{figure}

\begin{figure}
\includegraphics[width=60mm,angle=-90,trim= 0 15 10 38,clip]{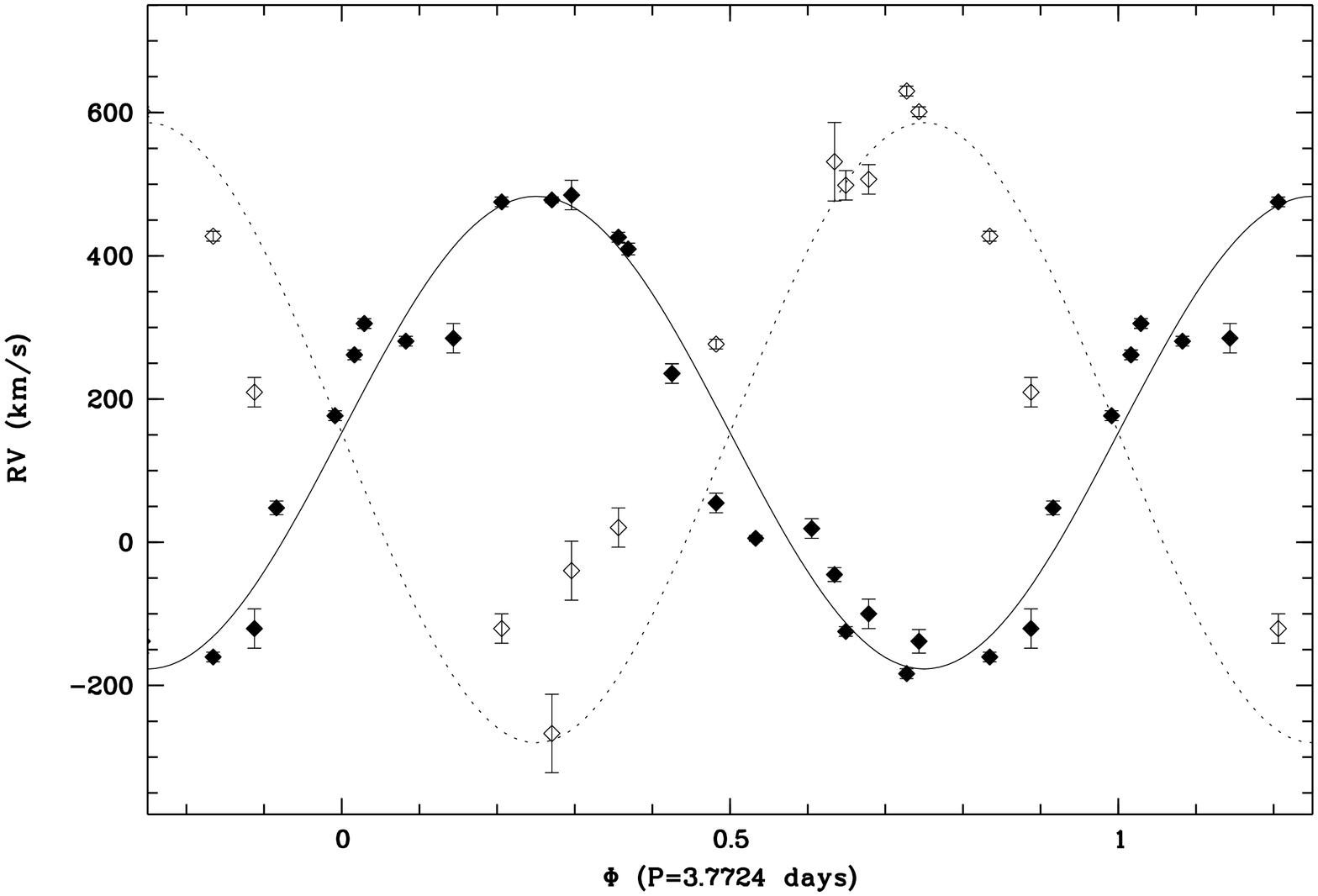}
\includegraphics[width=60mm,angle=-90,trim= 0 15 10 38,clip]{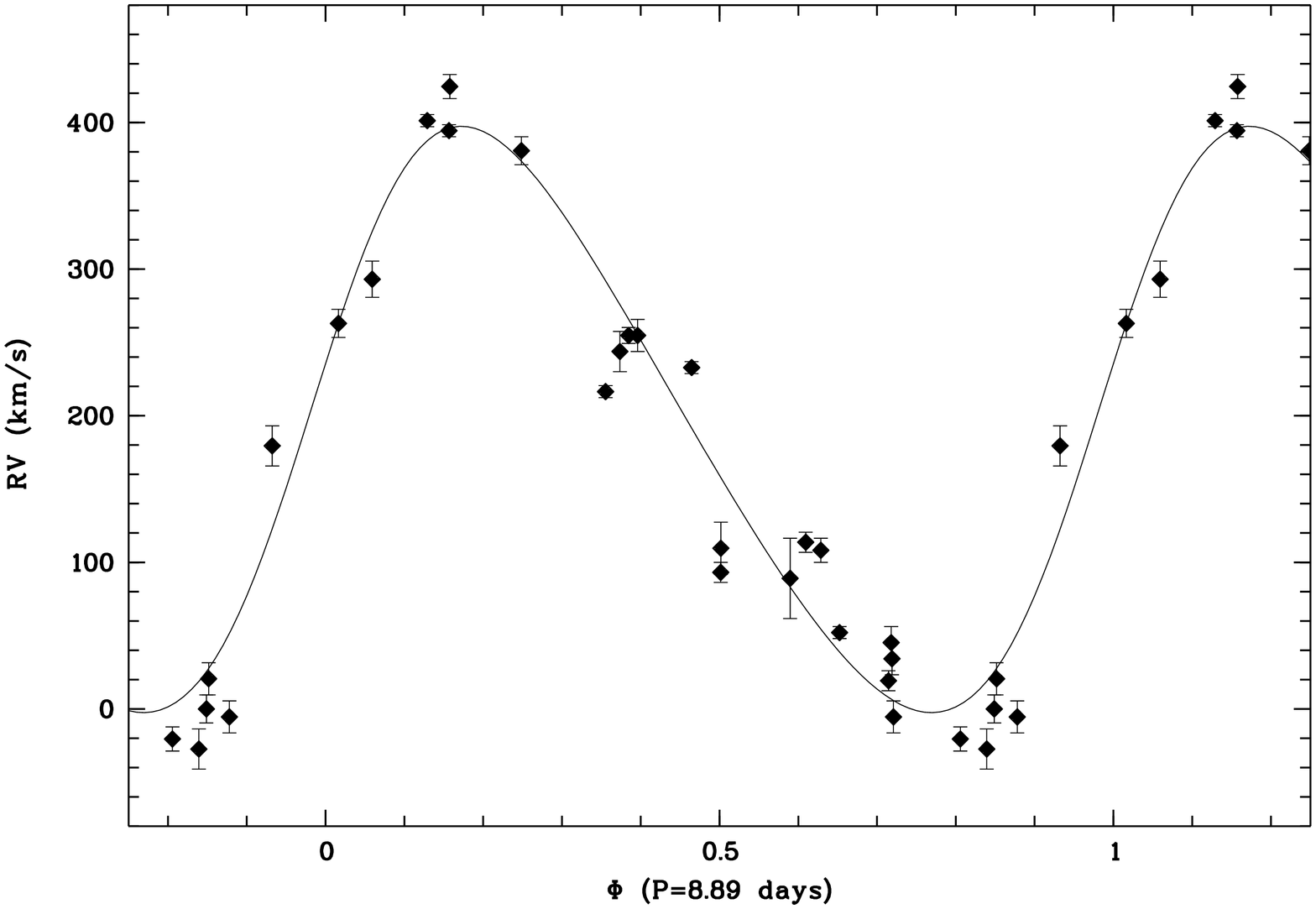}
\includegraphics[width=60mm,angle=-90,trim= 0 15 10 38,clip]{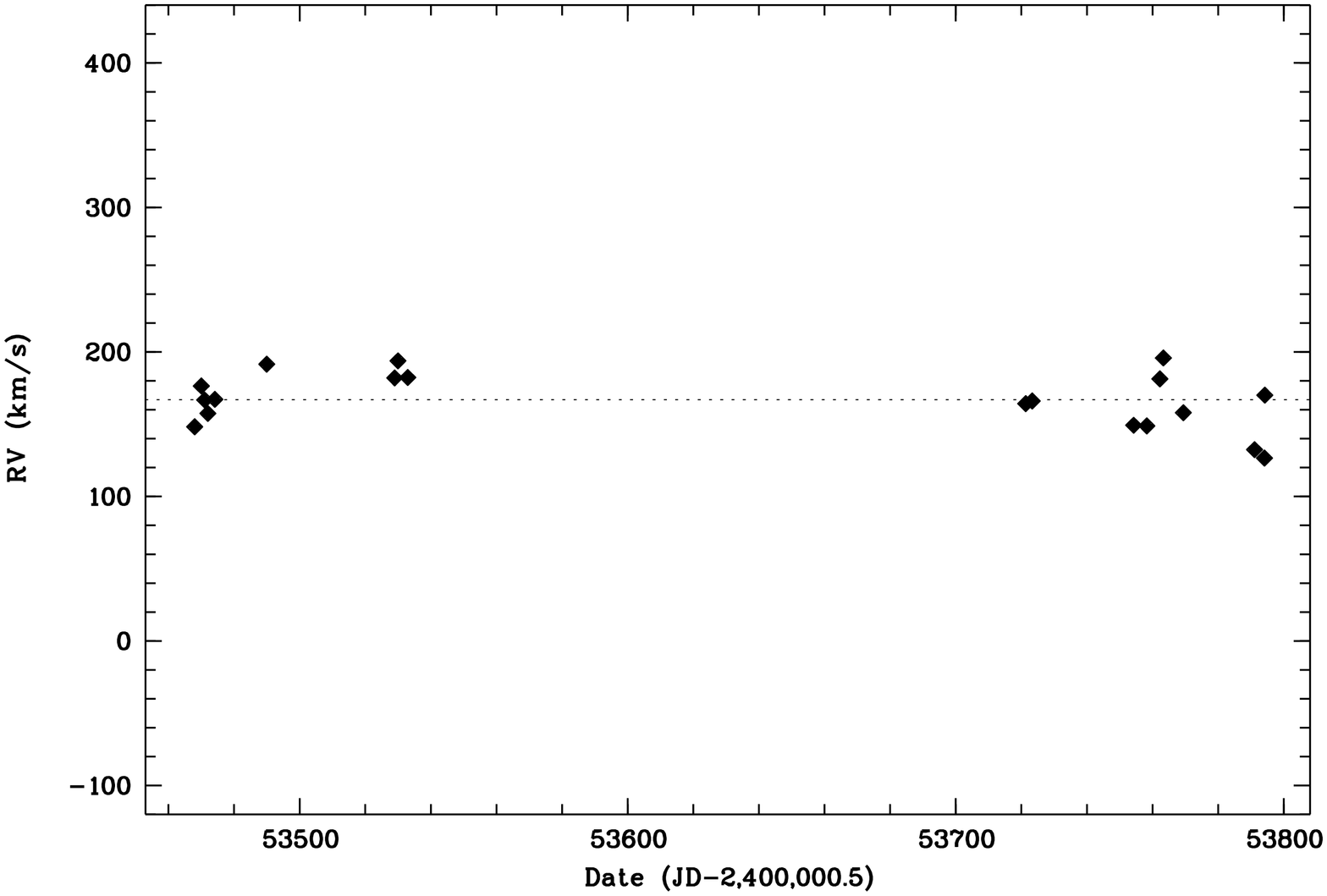}
\caption{Orbital solutions obtained from weighted fits for A1 (upper
panel) and C (middle panel), folded in the respective phases. In A1,
the primary is indicated by filled symbols and the solid curve, while
the secondary is shown with open symbols and the dotted curve. In star
C, only the primary is visible. RVs of star B are constant during the
observations (lower panel, same scale as for star C) with a scatter of
$\sigma = 20$ kms$^{-1}$ . The dashed line indicates the systemic
velocity of the star B, $\gamma = (167 \pm 6)$ kms$^{-1}$.}
\label{orbits}
\end{figure}

Orbital solutions for A1 and C were fitted to the RVs using
E\textsc{lem} (\citealt{Marchenko94}) and applying $1/\sigma_{\rm
RV}^{2}$ weights to the RVs. For star A1, the period reported by
\citet{Moff04}, $P=3.7724$ d, was adopted, and for both A1 and C the
orbital periods were fixed. In A1, RVs of both the primary and the
secondary were fitted simultaneously, forcing a circular orbit and the
systemic velocity to be the same for both components. While emission
lines in WR stars are well known to display some degree of redshift
with respect to the true systemic velocity of the star (see
e.g. \citealt{S08}), both WR components in A1 will display the same
amount of redshift. Our approach is further justified by the fact that
all three stars A1, B, and C show comparable systemic velocities.

Due to the low quality of the secondary's RV determination
during the ``blue'' half-wave of its orbital motion, the fit of the
secondary's orbit is dominated by the ``red'' half-wave. This leads to
a large RV amplitude for the secondary and, by consequence, a large
mass ratio. A significantly smaller RV amplitude would have been
obtained if the systemic velocity of the secondary were left as a free
fitting parameter. However, given the problems with the `blue
half-wave described above, we believe that it is more sensible to rely
on the red half-wave with the smaller error bars and fix the systemic
velocity.


Orbital fits for A1 and C are shown in Figure \ref{orbits}; the
respective orbital parameters are given in Tables \ref{orparamsA1} and
\ref{orparamsC}. Combining the orbital parameters of both components
of A1 with the inclination angle obtained by \citet{Moff04}, $i =
71^{\circ}$, and using the usual relations, we obtain absolute masses
of $M_{\rm 1} = (116 \pm 31)$ $M_{\sun}$ for the primary and $M_{\rm
2} = (89 \pm 16)$ $M_{\sun}$ for the secondary component of A1,
respectively. Due to the problems with the secondary's RVs described
above, the uncertainties on the orbits, expressed by the large
$\sigma_{\rm o-c}$, and hence on the masses are uncomfortably
large. However, \citet{Drissen95} have reported an absolute magnitude
of $M_{\rm V} = -7.5$ for A1, i.e. brighter by $\sim0.5$ mag than
WR20a, for which \citet{Rauw07} find $M_{\rm V} = 7.04 \pm
0.25$. Since A1 and WR20a have very similar spectral types and, thus,
bolometric corrections, A1's brighter magnitude directly translates
into a larger bolometric luminosity, qualitatively consistent with our
result that A1 is more massive than WR20a. However, better data are
currently being taken to verify the results.

\begin{table}
\caption{Orbital parameters for both the primary and the secondary
component of A1 from the combined, weighted fit, forcing a circular
solution. The inclination angle and the period have been adopted from
Moffat et al. (2004).}
\label{orparamsA1}
\begin{tabular}{l r@{}@{ }p{2mm}@{ }@{}l r@{}@{ }p{2mm}@{ }@{}l r@{}@{ }p{2mm}@{ }@{}l}
\hline
Parameter                  &  \multicolumn{3}{c}{Primary} & \multicolumn{3}{c}{}  & \multicolumn{3}{c}{Secondary} \\
\hline
$P$ [days]                 &      &       &     &          & 3.7724 &       &      &       &    \\
$i$ [$^{\circ}$]           &      &       &     &          &   71   &       &      &       &    \\
$e$                        &      &       &     &          &    0   &       &      &       &    \\
$E_{0}$ [2,450,000.5+]   &      &       &     & 3765.25 &  $\pm$ & 0.03  &      &       &    \\
$\gamma$ [kms$^{-1}$]      &      &       &     &      153 &  $\pm$ &  12   &      &       &    \\
$K$ [kms$^{-1}$]           & 330  & $\pm$ & 20  &          &        &       &  433 & $\pm$ & 53 \\
$\sigma_{\rm o-c}$ [kms$^{-1}$] & &   42  &     &          &        &       &      &  82   &  \\
$M$ [$M_{\odot}$]          & 116  & $\pm$ & 31  &          &        &       &   89 & $\pm$ & 16 \\
\hline
\end{tabular}
\end{table}

\begin{table}
\centering
\caption{Orbital parameters for the WN6ha (primary) component in star
C.}
\label{orparamsC}
\begin{tabular}{l r@{}@{ }p{2mm}@{ }@{}l}
\hline
Parameter & \multicolumn{3}{c}{Primary} \\
\hline
$P$ [days]                        & 8.89 &  $\pm$ & 0.01             \\
$e$                               & 0.30 &  $\pm$ & 0.04             \\
$\omega$ [$^{\circ}$]             &  281 &  $\pm$ & 7        \\
$T_{0}$ [2,450,000.5+]        & 3,546.61 &  $\pm$ & 0.18 \\
$\gamma$ [kms$^{-1}$]             & 186  &  $\pm$ & 6        \\
$K$ [kms$^{-1}$]                  & 200  &  $\pm$ & 23        \\
$\sigma_{\rm o-c}$ [kms$^{-1}$]   &      &  27    &   \\ 

\hline
\end{tabular}
\end{table}


In star C, the secondary is not visible in the spectrum; hence, we
temporarily classify the system as SB1. While the intrinsic brightness
difference between A1 and C is $\sim0.5$ mag (\citealt{CroDess98}), it
is well possible that the WN6ha component in C is as massive as the
primary in A1, where the luminosities of the primary and secondary are
more similar; for the same reason, a large mass ratio can be expected
between the primary and secondary component in C.

Star B, on the other hand, displays neither significant RV nor
photometric variations (cf. \citealt{Moff04}) nor a strong X-ray flux.
The latter is a strong indication against B's possible long-period
binarity. Therefore, B most likely is a truly single star. Remarkably,
\citet{CroDess98} report a luminosity for B which is only very
slightly lower than that of the (unresolved) system A1. This renders B
significantly more massive than the primaries in both A1 and C, and
puts it close to (or even in excess of) the putative IMF cut-off mass,
unless B turns out to be an unresolved, long-period binary as well.
Clearly, line-blanketed atmosphere models will provide valuable
insights into this problem once the known, very massive binary systems
such as A1 and WR20a have been used as yardsticks.

\section{Summary and Conclusion}

We have obtained repeated, spatially resolved, AO assisted, near-IR
spectroscopy of the three central WN6ha stars in HD 97950, the core of
the young, unevolved and very massive Galactic cluster NGC3603. One of
the stars, A1, is a previously known, double-eclipsing binary with an
orbital period of 3.77 days (\citealt{MoffNiem84}; \citealt{Moff85};
\citealt{Moff04}), while a second star, C, was newly identified as a
binary in the present study. The third star, B, showed constant RVs
over the observed time interval, and therefore is most likely not a
binary, which is in line with its normal X-ray luminosity
(\citealt{Moff02}).

While in star C only the primary (WN6ha) component is visible -- the
system is therefore classified as an SB1 binary --, A1 consists of two
emission-line stars, most likely of similar, but hot identical
spectral types. From the radial-velocity curves of the two components
and the known inclination angle of the system (\citealt{Moff04}), we
derived component masses of $M_{\rm 1} = (116 \pm 31) M_{\sun}$ for the
primary and $M_{\rm 2} = (89 \pm 16) M_{\sun}$ for the secondary,
respectively. Despite the large uncertainties, we consider the primary
WN6ha component of A1 to be the most massive star ever directly
weighed.

While the primary component of C might have a mass similar to or even
greater than that of A1's primary, it is possible that star B, single
yet only slightly fainter than the combined binary system A1, is indeed
the most massive member in NGC3603 and, therefore, the most massive
main-sequence star known in the Galaxy.


\label{lastpage}

\end{document}